\documentclass[%
preprint, 
superscriptaddress,
showpacs,
showpacs,preprintnumbers,
amsmath,amssymb,
aps,
]{revtex4-1}

\usepackage[dvipdfmx]{graphicx}
\usepackage{dcolumn}
\usepackage{bm}
\usepackage{multirow}
\usepackage[mathlines]{lineno}
\usepackage{comment}

\usepackage{xr}
\usepackage{color}

\begin{document}

\preprint{APS/123-QED}

\title{Phonon-mediated superconductivity in the Sb-square-net compound LaCuSb$_2$}

\author{Kazuto~Akiba}
\email{akb@okayama-u.ac.jp}
\affiliation{
Graduate School of Natural Science and Technology, 
Okayama University, Okayama 700-8530, Japan
}

\author{Tatsuo~C.~Kobayashi}
\affiliation{
Graduate School of Natural Science and Technology, 
Okayama University, Okayama 700-8530, Japan
}

\date{\today}

\begin{abstract}
We investigated the electronic structure and superconducting properties of 
single-crystalline LaCuSb$_2$.
The resistivity, magnetization, and specific heat measurements showed that
LaCuSb$_2$ is a bulk superconductor. 
The observed Shubnikov-de Haas oscillation and magnetic field dependence of the Hall resistivity can be reasonably understood assuming a slightly hole-doped Fermi surface.
Electron-phonon coupling calculation clarified the difference from the isostructural compound LaAgSb$_2$ indicating that (i) low-frequency vibration modes related to the interstitial layer sandwiched between the Sb-square nets significantly contribute to the superconductivity and (ii) carriers with sizable electron--phonon coupling distribute isotropically on the Fermi surface.
These are assumed to be the origin of higher superconducting transition temperature compared with LaAgSb$_2$.
We conclude that the superconducting properties of LaCuSb$_2$ can be understood within the framework of conventional phonon-mediated mechanism.

\end{abstract}

\maketitle

\section{Introduction}
\label{sec_intro}
The relationship between superconductivity (SC) and other ordered state
can provide new insights into the mechanism of the SC.
For example, the numbers of the so-called unconventional superconductors,
whose properties cannot be understood based on the conventional phonon-mediated mechanism, have been discovered adjacent to an ordered state in the phase diagram.
Heavy fermion systems allow the exploration of unconventional SC
near the quantum critical point (QCP),
at which the spin fluctuation effect plays an important role in the physical properties \cite{Moriya_SCR}.
As the SC emerges only in the vicinity of the QCP,
the phase diagram represents a characteristic dome-like shape \cite{Movshovich_1996, Grosche_1996, Mathur_1998}.
This fact indicates that spin fluctuation plays an important role in the pairing mechanism of this material class.

Another example is a critical point of charge order.
Recent studies reported an emergence of the SC or notable enhancement
of the transition temperature ($T_c$)
near the critical point of a charge density wave (CDW) \cite{Morosan_2005, Kusmartseva_2009, Gruner_2017, Chen_2021, Fu_2021}.
Although a possible existence of quantum fluctuation
associated with the CDW has been discussed,
the origin of the enhancement of SC remains unclear.
To elucidate whether there is an unconventional pairing mechanism around the CDW critical point,
model materials are necessary to investigate systematically the relationship between SC and CDW.

As a candidate material,
the authors have recently focused on an intermetallic compound LaAgSb$_2$.
LaAgSb$_2$ crystallizes a tetragonal structure (space group $P4/nmm$, \#129) with a Sb-square net structure
\cite{Brylak_1995}.
This material class has attracted attention as an extending system possessing Dirac fermion
for the band folding of a $4^4$-square net \cite{Hoffmann_1987, Klemenz_2019, Klemenz_2020}.
LaAgSb$_2$ shows successive CDW transitions at $T_{CDW1}\sim 210$ K and $T_{CDW2}\sim 190$ K at ambient pressure \cite{Myers_1999a, Song_2003}, which can be systematically suppressed by applying hydrostatic pressure \cite{Budko_2006, Torikachvili_2007}.
The authors investigated the transport properties of LaAgSb$_2$ at high pressures
and established the phase diagram and the Fermi surface (FS) subjected to pressure \cite{Akiba_2021, Akiba_2022}.
Further, we discovered SC with $T_c\sim 0.3$ K coexisting with the CDWs
at ambient pressure.
The $T_c$ was considerably enhanced up to 1 K only around the critical pressure of CDW1 \cite{Akiba_2022_SC}.
Theoretical $T_c$ assuming the conventional phonon-mediated mechanism cannot reproduce the 1 K-order $T_c$ and its significant suppression in the normal metallic phase above 3.2 GPa,
which indicates that an additional mechanism is activated at the CDW critical point to reinforce the pairing interaction.
In addition, the electron--phonon coupling (EPC) calculation has indicated that the FSs derived from $p_x$ and $p_y$ orbitals of the Sb-square net
introduce a primary contribution to SC,
whereas the contributions from the rest of the FSs are considerably smaller. 
This indicates that the Sb-square-net structure is important
not only for the emergence of linear dispersion and the nesting of CDW1,
but also the primary conduction layer of SC.
Thus, LaAgSb$_2$ is a promising candidate used to elucidate the relationship between these orders.
To understand the origin of the enhancement of $T_c$,
careful comparison with related materials is of primary importance.

LaCuSb$_2$, the target of the present study,
crystallizes the identical structure with LaAgSb$_2$,
which is shown in Fig. \ref{fig01}(a).
In contrast with the LaAgSb$_2$ and another isostructural compound LaAuSb$_2$
\cite{Kuo_2019, Xiang_2020, Du_2020, Lingannan_2021},
no CDW transition is reported at ambient pressure,
and only SC at $T_c\sim 0.9$ K has been reported in polycrystalline samples by Muro \textit{et al}. \cite{Muro_1997}.
On the other hand,
a recent study investigated the band structure by angle-resolved photoemission spectroscopy and the magneto-transport properties on a single crystal \cite{Chamorro_2019}.
In this study, Chamorro \textit{et al.} experimentally reported
the Dirac-like linear dispersion,
possible weak antilocalization effect due to the two-dimensionality of the electronic structure,
linear magnetoresistance effect,
and Shubnikov-de Haas (SdH) oscillation with light cyclotron effective mass ($m_c^*$).
Based on the above, the authors suggested a possible realization of topologically nontrivial electronic states in LaCuSb$_2$.
Contrary to the previous study by Muro \textit{et al.},
however, the SC has not been observed.
Thus, whether the SC is intrinsic or not has not been conclusive.

From the viewpoint of the analogy with LaAgSb$_2$ and LaAuSb$_2$,
it is worth noting that the unit cell volume of LaCuSb$_2$ is 198.1 \AA$^3$
($a=4.3690$ \AA, $c=10.376$ \AA) \cite{Sologub_1994},
which is significantly smaller compared with 209.84 \AA$^3$ of LaAgSb$_2$
($a=4.3941$ \AA, $c=10.868$ \AA) \cite{Akiba_2021},
and 205.80 \AA$^3$ of LaAuSb$_2$
($a=4.441$ \AA, $c=10.435$ \AA) \cite{Lingannan_2021}.
This suggests that LaCuSb$_2$ can be regarded as the pressurized counterpart of
LaAgSb$_2$ and LaAuSb$_2$.
Thus, elucidating the physical properties of LaCuSb$_2$ would contribute to the understanding of the La-based intermetallic system under pressure.

In this study, we investigated the electronic structure and superconducting properties of LaCuSb$_2$
using experimental and computational techniques.
We observed $T_c$ at 1.0 K in electrical resistivity, magnetization,
and specific heat measurements,
certifying that SC is a bulk effect.
In magneto-transport measurements,
we revealed field-angular dependence of the SdH oscillation and positive Hall resistivity.
First-principles calculation showed that
the experimental results are consistently understood by a hole-doped FS.
In contrast to the case of LaAgSb$_2$,
EPC calculations showed that
(i) several low-frequency phonon modes related to the interstitial layer sandwiched between the Sb-square-net layers
shows significant EPC and 
(ii) the momentum-resolved EPC distributes isotropically over the entire FS.
The factors listed above doubled the integrated EPC strength compared with that of LaAgSb$_2$,
contributing to the higher $T_c$ of LaCuSb$_2$.
The experimental $T_c$ is reasonably reproduced by the McMillan--Allen--Dynes formalism;
thus, we conclude that the SC of LaCuSb$_2$ is reasonably understood based on the conventional phonon-mediated mechanism without considering possible quantum critical phenomena.

\section{Experimental method}
\label{sec_exp_method}

\begin{figure}[]
\centering
\includegraphics[]{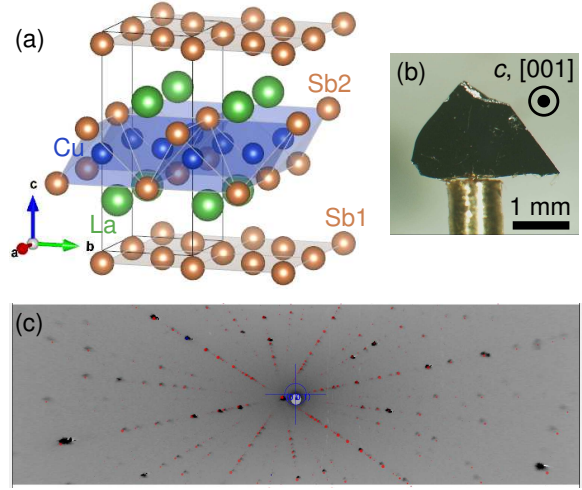}
\caption{
(a) Crystal structure of LaCuSb$_2$.
(b) Photograph of a single crystal.
(c) Back-reflection Laue pattern with X-rays applied parallel to the [001] direction. 
Red points represent the simulation (see main text).
\label{fig01}}
\end{figure}

Single crystals of LaCuSb$_2$ were obtained by the Sb self-flux method.
La (99.9\%), Cu (99.99\%), and Sb (99.9999\%) with a molar ratio of 1:2:20 were placed in an alumina crucible and sealed in a quartz ampoule with argon gas.
After the mixture was heated to $1150\ {}^\circ\mathrm{C}$, it dwelled at this temperature for 12 h.
It was then cooled to $670\ {}^\circ\mathrm{C}$ for 120 h.
The flux was removed using a centrifuge separator.
We obtained plate-like crystals with millimeter-size dimensions, as shown in Fig. \ref{fig01}(b).
The obtained samples were investigated by Laue diffraction measurements using IPX-YGR
(IPX Co., Ltd.) based on a back-reflection configuration.
Clear Laue spots were observed as shown in Fig. \ref{fig01}(c), ensuring the high quality of the single crystal.
The pattern was reproduced by simulations (red points) executed using QLaue \cite{qlaue_url}
assuming the reported lattice constants \cite{Sologub_1994}.

The resistivity measurements were performed following a standard four-terminal method.
We utilized a Model 370 AC resistance bridge (Lake Shore Cryotronics, Inc.)
or a combined system of 2400 sourcemeter and 2182A nanovoltmeter (Keithley Instruments). 

Magnetization measurements were performed using
a direct current superconducting quantum interference device
(dc-SQUID, Tristan Technologies, Inc.).
A signal-pick-up coil wound using a superconducting NbTi wire was
connected to the sensing terminals on the dc-SQUID.
The dc-SQUID was placed in a He bath and its temperature was held at 4.2 K.
The signal from the dc-SQUID was acquired with a controller via a communication cable,
and we finally obtained the voltage signal, which was proportional to the change in the magnetic flux inside the pickup coil.
In the magnetization measurement, we did not intentionally apply an external magnetic field.
The measurements were conducted in a residual geomagnetic field.

Specific heat was measured by the relaxation method using a homemade calorimetry cell.
A calibrated thermometer (2 k$\Omega$-RuO$_2$ chip resistor, KOA Corporation)
and a heater (120 $\Omega$-strain gauge, Kyowa Electronic Instruments Co., Ltd.)
were attached to the sample stage (a piece of Ag ribbon) with stycast 1266.
The sample stage was suspended in the vacuum space by manganin wires, which served both as thermal leak paths and current leads.
The temperature dependence of the addenda heat capacity and
the thermal conductivity of the manganin were determined beforehand by a measurement without sample.
The sample was attached to the stage with Apiezon N grease.
We measured several relaxation processes and obtained the temperature dependence of total heat capacity, which contained both sample and addenda contributions.
We finally obtained the sample heat capacity by subtracting the known addenda heat capacity.

The details of magnetization and specific heat measurements are described in
Supplemental Material of \cite{Akiba_2022_SC}.

Temperatures down to 2 K were realized by using a He-gas-flow-type optical cryostat (Oxford Instruments).
Temperatures down to 50 mK were realized using a homemade $^3$He/$^4$He dilution refrigerator.
The magneto-transport properties in the presence of magnetic fields were measured using a superconducting magnet with a variable-temperature insert (Oxford Instruments, $B < 8$ T and $T > 1.6$ K).
The field-angular dependence of the resistivity was measured using a homemade mechanical rotator, which can uniaxially rotate the stage in the variable-temperature insert.

\section{Computational method}
\label{sec_comp_method}

The structural optimization and band structure calculation based on the density-functional theory (DFT)
were performed using the Quantum ESPRESSO (QE) package \cite{Giannozzi_2009, Giannozzi_2017}.
We employed scalar-relativistic projector-augmented wave pseudopotentials with the Perdew--Burke--Ernzerhof exchange-correlation functional \cite{pseudo}.
We used a cutoff of 95 Ry and 950 Ry for the plane-wave expansions of
wave functions and charge density, respectively,
and a $\Gamma$-shifted Monkhorst--Pack $10\times10\times6$ $\bm{k}$-point grid for the self-consistent calculation.
Self-consistent calculations were performed with a threshold of $1.0\times 10^{-8}$ Ry.
Structural optimization was performed using convergence thresholds of $1.0\times 10^{-5}$ Ry for the total energy change and $1.0\times 10^{-4}$ Ry/Bohr for the forces.
Fully relaxed lattice constants and positions for La (0.25, 0.25, $z_{La}$) and Sb2 (0.75, 0.75, $z_{Sb2}$) are shown in Table \ref{tab_opt}.
The obtained lattice constants are close to the experimental values,
$a=4.3690$ \AA, $c=10.376$ \AA \cite{Sologub_1994},
and the atomic coordinates agree with the previous calculation, $z_{La}=0.2537$
and $z_{Sb2}=0.3535$ \cite{Ruszala_2018}.
The crystal structure was visualized by VESTA \cite{Momma_2011}.

\begin{table}[]
\caption{\label{tab_opt}
Lattice constants $a$ and $c$ and atomic coordinates for La (0.25, 0.25, $z_{La}$) and Sb2 (0.75, 0.75, $z_{Sb2}$) obtained based on the structural optimization.
}
\begin{ruledtabular}
\begin{tabular}{lllll}
space group&$a$ (\AA) & $c$ (\AA) & $z_{La}$ & $z_{Sb2}$ \\
\hline
$P4/nmm$ (\#129)&4.423 & 10.293 &0.2529&0.3524
\end{tabular}
\end{ruledtabular}
\end{table}

Based on the DFT calculation,
we constructed the tight-binding Hamiltonian using Wannier90 \cite{Pizzi_2020}.
We confirmed that 16 Wannier orbitals
(La-$d_{z^2}$, La-$d_{x^2-y^2}$, and Sb-$p$)
are sufficient to reproduce the DFT band structure at the Fermi level.
Particularly, in the band structure calculation with orbital character projections (Fig. \ref{fig08}),
we employed 46 Wannier orbitals
(La-$d$, La-$f$, Ag-$d$, and Sb-$p$)
to represent the band character accurately.
Visualization of the Wannier-interpolated FS (Fig. \ref{fig06}) was performed using the FermiSurfer \cite{Kawamura_2019}.
The simulations of quantum oscillation frequency $F$ and
cyclotron effective mass $m_c^*$
(Fig. \ref{fig06} and Table \ref{tab_qo})
were performed using the SKEAF code \cite{Rourke_2012}.

The simulation of electrical conductivity tensor $\bm\sigma$ (Fig. \ref{fig07}) was conducted based on the Boltzmann equation
within the relaxation-time approximation using WannierTools \cite{Wu_2017, Zhang_2019}.
In the framework described above, the conductivity tensor was represented by
\begin{equation}
\sigma_{ij}^{(n)}=\dfrac{e^2}{4\pi^3}\int d\bm k v_i^{(n)}(\bm k) \tau_n \bar{v}_j^{(n)}(\bm k)\left(-\dfrac{\partial f_{FD}}{\partial \epsilon}\right)_{\epsilon=\epsilon_n(\bm k)}.
\label{eq_CF}
\end{equation}
Herein, $e$, $f_{FD}$, and $n$ represent the elemental charge, Fermi--Dirac distribution function, and band index, respectively.
$\tau_n$ represents the relaxation time of the $n$th band, which is assumed to be independent of $\bm k$.
Because of the energy derivative of the Fermi--Dirac distribution function,
$\sigma_{ij}^{(n)}$ was determined by the states within the thermal energy width of $\sim k_B T$ near the Fermi level.
We set $T=10$ K to define the thermal energy width. 
$\bm v^{(n)}(\bm k)$ represents the velocity defined by the gradient of the energy in the reciprocal space as
\begin{equation}
\bm v^{(n)}(\bm k)=\dfrac{1}{\hbar}\dfrac{\partial \epsilon_n(\bm k)}{\partial \bm k}.
\end{equation}
$\bar{\bm v}^{(n)}(\bm k)$ represents the weighted average of velocity over the orbit, which is defined as
\begin{equation}
\bar{\bm v}^{(n)}(\bm k)=\int_{-\infty}^{0}\dfrac{dt}{\tau_n}e^{t/\tau_n}\bm v^{(n)}[\bm k(t)].
\label{eq_wv}
\end{equation}
The historical motion of $\bm k(t)$ under a magnetic field $\bm B$ was obtained by the equation of motion
\begin{equation}
\dfrac{d\bm k(t)}{dt}=-\dfrac{e}{\hbar}\bm v^{(n)}[\bm k(t)]\times \bm B,
\end{equation}
where $\bm k(t=0)=\bm k$.
We adopted a $101^3$ $\bm k$ mesh
for the integration over the first Brillouin zone.

The phonon calculations were performed based on the density functional perturbation theory (DFPT) with the optimized tetrahedron method \cite{Kawamura_2014} implemented in QE.
A convergence threshold of $1.0\times 10^{-14}$ Ry was employed for the DFPT self-consistent iterations.
The phonon dispersions were calculated using a $\Gamma$-centered $4^3$ $\bm{q}$-point grid.
The lattice vibrations were visualized using the phonon website \cite{phononwebsite_url}.

EPC properties (Fig. \ref{fig09})
were calculated using the EPW \cite{Ponce_2016} code.
The electron-phonon matrix element,
which describes the scattering process (from band $n$ to $m$) of an electron (wavenumber $\bm{k}$) by a phonon (wavenumber $\bm{q}$ and mode index $\nu$), is defined as
\begin{equation}
g_{mn, \nu}(\bm{k}, \bm{q})=\sqrt{\dfrac{\hbar}{2M \omega_{\bm{q}\nu}}}\langle \Psi_{m\bm{k}+\bm{q}}|\partial_{\bm{q}\nu}V|\Psi_{n\bm{k}}\rangle.
\end{equation}
where $M$ and $\hbar$ are the mass of the nuclei and reduced Planck constant, respectively.
$\omega_{\bm{q}\nu}$ represents the frequency of phonon with a wavevector $\bm{q}$ and mode $\nu$. 
$|\Psi_{n\bm{k}}\rangle$ is the electronic wavefunction for band index $n$ and wavevector $\bm{k}$
with an eigenvalue of $\epsilon_{n\bm{k}}$.
$\partial_{\bm{q}\nu}V$ is the derivative of the self-consistent potential associated with a phonon with a wavevector $\bm{q}$ and mode $\nu$.
Using $g_{mn, \nu}(\bm{k}, \bm{q})$, 
the phonon linewidth $\gamma_{\bm{q}\nu}$ and EPC strength $\lambda_{\bm{q}\nu}$ are represented as
\begin{equation}
\gamma_{\bm{q}\nu}=2\pi \omega_{\bm{q}\nu}\sum_{nm}\int_{BZ}\dfrac{d\bm{k}}{\Omega_{BZ}} |g_{mn, \nu}(\bm{k}, \bm{q})|^2\delta(\epsilon_{n\bm{k}}-\epsilon_F)\delta(\epsilon_{m\bm{k}+\bm{q}}-\epsilon_F),
\label{eq_linewidth_int}
\end{equation}
\begin{equation}
\lambda_{\bm{q}\nu} = \dfrac{2}{N(\epsilon_F)\omega_{\bm{q}\nu}} \sum_{nm}\int_{BZ}\dfrac{d\bm{k}}{\Omega_{BZ}} |g_{mn, \nu}(\bm{k}, \bm{q})|^2\delta(\epsilon_{n\bm{k}}-\epsilon_F)\delta(\epsilon_{m\bm{k}+\bm{q}}-\epsilon_F)=\dfrac{\gamma_{\bm{q}\nu}}{\pi N(\epsilon_F)\omega_{\bm{q}\nu}^2}.
\label{eq_lambda_int}
\end{equation}
where $N(\epsilon_F)$ is the density of states per spin at the Fermi level $\epsilon_F$. The integral was evaluated over the Brillouin zone with an $\Omega_{BZ}$ volume.
$\delta(\epsilon)$ represents the Dirac delta function.
Alternatively, we can represent the EPC strength
in the $\bm{k}$-space as
\begin{equation}
\lambda_{n\bm{k}}=\sum_{m\nu}\int_{BZ}\dfrac{d\bm{q}}{\Omega_{BZ}}\dfrac{2}{\omega_{\bm{q}\nu}}|g_{mn, \nu}(\bm{k}, \bm{q})|^2\delta(\epsilon_{m\bm{k}+\bm{q}}-\epsilon_F).
\end{equation}
Hereafter, we omit the band index $n$ and simply write it as $\lambda_{\bm{k}}$.

Using $\lambda_{\bm{q}\nu}$, the Eliashberg spectral function $\alpha^2F(\omega)$ can be obtained by calculating its integrated value over the Brillouin zone as follows,
\begin{equation}
\alpha^2F(\omega) = \dfrac{1}{2}\sum_{\nu}\int_{BZ}\dfrac{d\bm{q}}{\Omega_{BZ}} \omega_{\bm{q}\nu} \lambda_{\bm{q}\nu} \delta(\omega-\omega_{\bm{q}\nu}).
\label{eq_a2f_int}
\end{equation}
We estimated the superconducting transition temperature $T_c^{MAD}$ using the McMillan--Allen--Dynes formula \cite{McMillan_1968, Dynes_1972, Allen_1975}
\begin{equation}
T_c^{MAD} = \dfrac{\omega_{log}}{1.2}\exp \left(-\dfrac{1.04(1+\lambda)}{\lambda-\mu_c^*(1+0.62\lambda)}\right).
\end{equation}
Herein, $\lambda$ is defined using the Eliashberg spectral function and maximum phonon frequency $\omega_{max}$ as
\begin{equation}
\lambda = 2\int_0^{\omega_{max}} d\omega \dfrac{\alpha^2F(\omega)}{\omega},
\end{equation}
and $\omega_{log}$ is a logarithmic average of the phonon frequency defined as
\begin{equation}
\omega_{log}=\exp\left( \dfrac{2}{\lambda}\int_0^{\omega_{max}} d\omega \ln\omega \dfrac{\alpha^2F(\omega)}{\omega}\right).
\end{equation}
$\mu_c^*$ represents the Coulomb pseudopotential,
which is treated as an empirical parameter to express the Coulomb interaction.
For typical metals, $\mu_c^*$ is known to take values around 0.1 \cite{Morel_1962}.
In the present study, we assumed $\mu_c^*=0.1$.

We used coarse $8^3$ $\bm{k}$ and $4^3$ $\bm{q}$ meshes
for the initial calculation of the electronic Hamiltonian, dynamical matrix,
and electron-phonon matrix.
To calculate the EPC properties on arbitrary, dense Brillouin zone grids,
an interpolation scheme described in \cite{Giustino_2007} was applied using EPW.
In this procedure, we used 16 Wannier orbitals.
The integrations over the Brillouin zone were performed 
on uniform $75^3$ $\bm{k}$ and $15^3$ $\bm{q}$ meshes.
The Dirac delta functions
were smeared with widths of 25 meV for electrons
and 0.05 meV for phonons.

\section{Results}
\label{sec_results}

\begin{figure}[]
\centering
\includegraphics[]{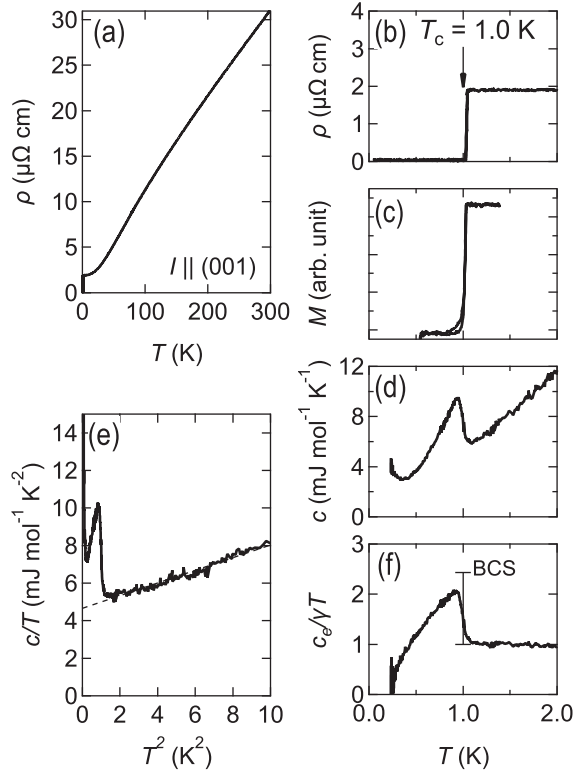}
\caption{
(a) Temperature dependence of resistivity $\rho$.
Superconducting transition at $T_c=1.0$ K in (b) $\rho$, (c) magnetization $M$, and
(d) specific heat $c$.
(e) Plot of $c/T$ as a function of $T^2$.
Extrapolation of the linear dependence to 0 K (broken line) yields the electron-specific heat coefficient $\gamma=4.65$mJ mol$^{-1}$ K$^{-2}$.
(f) Temperature dependence of $c_{e}/(\gamma T)$, where $c_{e}$ is obtained by subtracting nuclear and lattice contributions from $c$ (see main text).
The vertical scale represents the jump expected from the Bardeen--Cooper--Schrieffer (BCS) theory.
\label{fig02}}
\end{figure}

First, we show the temperature dependence of in-plane resistivity ($\rho$) in Fig. \ref{fig02}(a).
Metallic behavior without any indication of phase transition is exhibited,
and a zero-resistivity state emerges below 1.0 K
as seen in the low-temperature data in Fig. \ref{fig02}(b).
The onset of the resistivity drop is consistent with that of the polycrystalline sample \cite{Muro_1997}.
Accompanied by the resistivity drop,
the magnetization exhibits a distinct and abrupt anomaly, as shown in Fig. \ref{fig02}(c), which is ascribed to the Meissner effect.
Additionally, the thermodynamic evidence of the SC is depicted by a sudden change
in the molar specific heat ($c$) at 1.0 K
as shown in Fig. \ref{fig02}(d).
The results described above demonstrate that LaCuSb$_2$ is a bulk superconductor.

From the specific heat data, we can obtain quantitative information.
When the temperature ($T$) is sufficiently lower than the Debye temperature,
$c$ is represented as $c=\alpha/T^2+\beta T^3+\gamma T$,
where the first, second, and third terms represent the nuclear, phonon, and electron contributions, respectively.
From the $c/T$--$T^2$ plot shown in Fig. \ref{fig02}(e),
we can obtain the coefficients $\beta=338\pm 7 \mu$J mol$^{-1}$ K$^{-4}$ and $\gamma=4.65\pm 0.03$mJ mol$^{-1}$ K$^{-2}$.

Additionally, we can recognize an upturn below 0.4 K,
which is scaled by $\alpha/T^2$.
This is ascribed to the tail of the Schottky-type nuclear specific heat caused by an electric quadrupole splitting,
which has been observed in isostructural LaAgSb$_2$ \cite{Akiba_2022_SC}.
From curve fitting below 0.4 K,
we obtained the coefficient $\alpha=259.4\pm 0.5 \mu$J mol$^{-1}$ K.
We have confirmed that the $\alpha$ value is reasonably explained by
nuclear quadrupole resonance frequencies obtained by first-principles calculations.

Using the obtained coefficients $\alpha$ and $\beta$,
we calculated the electronic specific heat $c_{e}$
by subtracting the nuclear and phononic contributions from $c$.
Figure \ref{fig02}(f) shows the temperature dependence of $c_{e}$ divided by $\gamma T$.
The observed jump of $c_{e}/(\gamma T)$ at 1.0 K is 1.09.
Although this is slightly smaller than the value of 1.43 expected by Bardeen--Cooper--Schrieffer
(BCS) theory,
the accordance is assumed to be reasonable.

\begin{figure}[]
\centering
\includegraphics[]{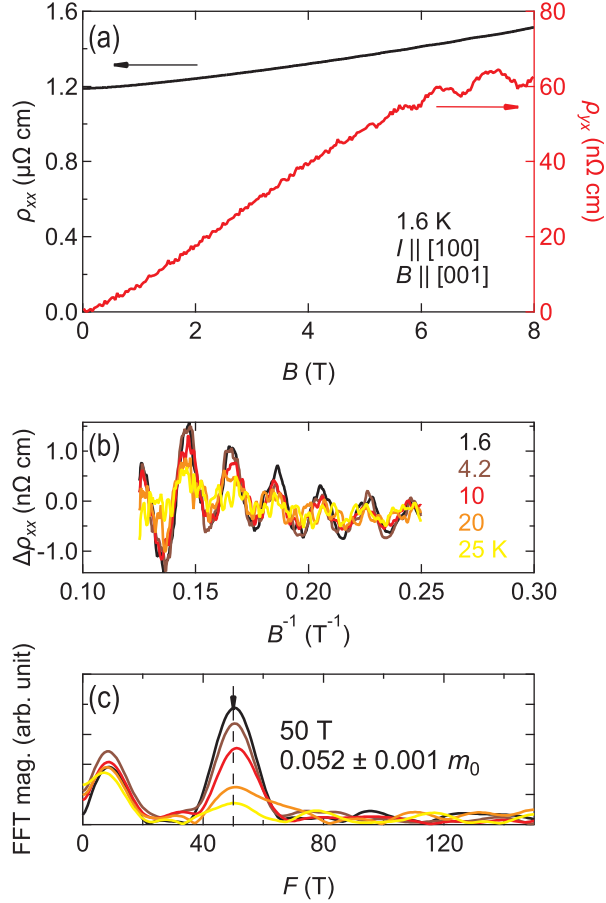}
\caption{
(a) Magnetoresistivity $\rho_{xx}$ (left axis) and Hall resistivity $\rho_{yx}$ (right axis) at 1.6 K.
(b) Oscillating component $\Delta \rho_{xx}$ superposed on $\rho_{xx}$
as a function of $B^{-1}$ at various temperatures.
(c) Fast Fourier transform (FFT) magnitude of $\Delta \rho_{xx}$ at the temperatures listed in (b).
Solid arrow indicates the peak at 50 T with cyclotron effective mass of 0.052 $m_0$. 
\label{fig03}}
\end{figure}

In the subsequent part, we describe the magneto-transport properties.
Figure \ref{fig03}(a) shows the in-plane magnetoresistivity ($\rho_{xx}$) and Hall resistivity ($\rho_{yx}$) at 1.6 K.
Herein, $B$ was applied along the [001] direction.
$\rho_{xx}$ shows weak positive magnetoresistance effect $\Delta \rho/\rho = [\rho_{xx}(B)-\rho_{xx}(0)]/\rho_{xx}(0)\sim 0.27$.
In this study, we did not observe a pronounced increase of $\rho_{xx}$ at the weak magnetic field reported previously \cite{Chamorro_2019},
which has been interpreted as a possible weak antilocalization effect.
$\rho_{yx}$ is positive up to 8 T, which indicates that the contribution of a hole-like orbit is dominant.
We note that a previous study reported negative $\rho_{yx}$ in similar conditions \cite{Chamorro_2019}, which apparently contradicts the present results.
We show subsequently that the positive $\rho_{yx}$ is supported by the FS geometry determined in the present study.
Additionally, $\rho_{yx}$ shows nonlinear behavior as a function of $B$,
as marked in the high-field region.
The possible origin of this nonlinearity will be discussed later.

We focus now on the SdH oscillations superimposed on $\rho_{xx}$.
Figure \ref{fig03}(b) shows the oscillating component $\Delta \rho_{xx}$ at various temperatures.
$\Delta \rho_{xx}$ was obtained by subtracting polynomial background from $\rho_{xx}$.
The amplitude systematically decreases as the temperature increases, which is expected from the conventional Lifshitz--Kosevich formula \cite{Shoenberg_MOM}.
From the fast Fourier transform (FFT) spectra shown in Fig. \ref{fig03}(c),
we identify a single frequency of $F = 50$ T with $m_c^*=0.052\pm0.001 m_0$,
where $m_0$ represents the bare mass of the electron.
The light $m_c^*$ is consistent with a previous report \cite{Chamorro_2019}.

\begin{figure}[]
\centering
\includegraphics[]{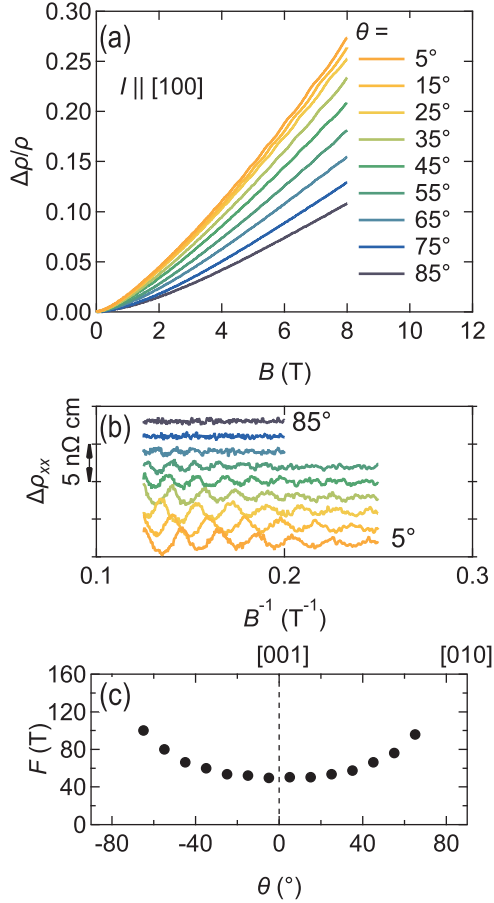}
\caption{
(a) Magnetoresistance effect $\Delta \rho/\rho$ at various magnetic field angles.
The angle $\theta$ is measured from [001].
(b) Oscillating component $\Delta \rho_{xx}$ at the $\theta$ shown in (a).
The data are vertically shifted for clarity.
(c) Field-angular dependence of the Shubnikov-de Haas (SdH) frequency $F$.
\label{fig04}}
\end{figure}

To obtain the geometrical information of the FS, we investigated the field-angular dependence of the SdH frequency.
Herein, the electric current $I$ was applied along [100],
and $B$ was rotated from [001] toward [010].
$B$ is always perpendicular to $I$ in this measurement.
We define $\theta$ as an angle measured from [001].
Figure \ref{fig04}(a) shows $\Delta \rho/\rho$ for various $\theta$.
The magnetoresistance effect is maximized (minimized) at $\theta \sim 0^{\circ}$ ($90^{\circ}$).
The SdH oscillation gradually becomes weak as $\theta$ increases,
and we cannot discern the oscillating component at $\theta > 65^{\circ}$.
The field-angular dependence of $F$ is summarized in Fig. \ref{fig04}(c),
which suggests that the observed FS has cylindrical or elongated ellipsoidal shapes.

\section{Discussion}
\label{sec_discussion}

\begin{figure}[]
\centering
\includegraphics[]{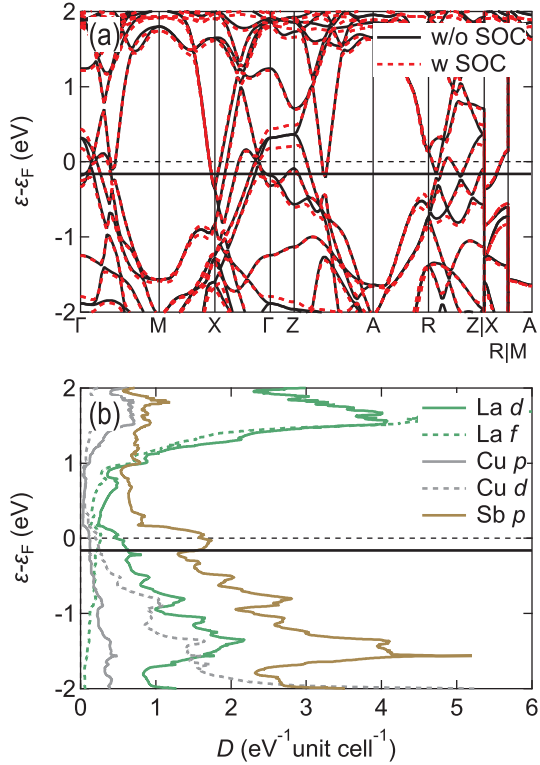}
\caption{
(a) Electronic band structure and (b) orbital-projected density of states $D$.
The energetic origin (horizontal broken line) is the Fermi level $\epsilon_F$.
The horizontal solid line indicates the hole-doped case to reproduce the experimental results (see main text).
The solid black and broken red band structures in (a) indicate the cases without and with spin-orbit coupling, respectively.
\label{fig05}}
\end{figure}

In the following parts,
we interpret our experimental results using first-principles calculations.
Figure \ref{fig05}(a) shows the band structure of LaCuSb$_2$.
The horizontal broken line ($\epsilon-\epsilon_F = 0$ eV) represents the Fermi level.
Herein, we can see that the spin-orbit coupling (SOC) can hardly modify the band structure at the Fermi level. Thus,
we adopted the band structure without SOC in the following discussion.
The band structure shares common features with that of isostructural LaAgSb$_2$,
which is shown in Supplemental Material \cite{SM_URL}.
As shown in Fig. \ref{fig05}(b), the major contribution to the density of states $D$ around the Fermi level comes from $p$-like orbital of Sb,
which corresponds to more than 50 \% of the total $D$. 

\begin{figure*}[]
\centering
\includegraphics[]{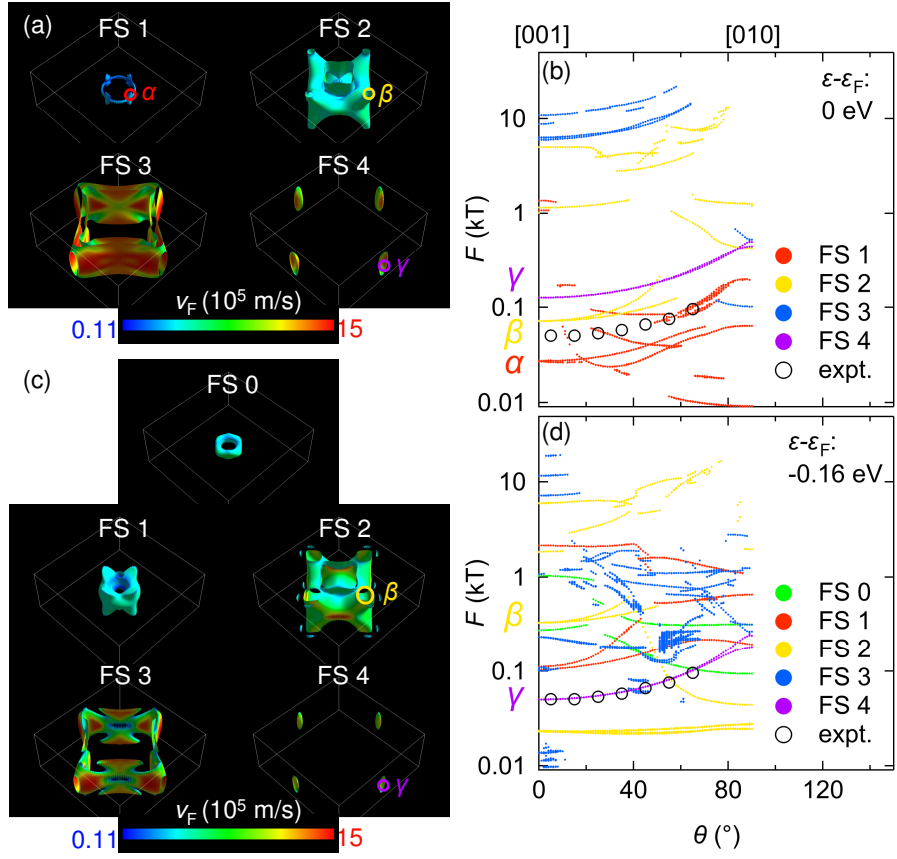 }
\caption{
(a) Fermi surface of the undoped case. (b) Calculated SdH frequencies based on the Fermi surface shown in (a).
(c) Fermi surface of the hole-doped case with a Fermi level shift of -0.16 eV. (b) Calculated SdH frequencies based on the Fermi surface shown in (c).
The large open markers in (b) and (d) represent the experimental (expt.) result.
\label{fig06}}
\end{figure*}

We attempted to determine the geometry of the FS based on the quantum oscillation.
Figure \ref{fig06}(a) shows the FS of LaCuSb$_2$,
in which the Fermi level is set at $\epsilon-\epsilon_F = 0$ eV in Fig. \ref{fig05}(a).
In the case of the aforementioned undoped condition, there are four FSs labeled FS 1--4:
the first two are holes, and the remaining two are electron surfaces.
FS 1 is tiny and can, thus, vanish by a slight upward shift of the Fermi level.
Although the overall features of FS 2-4 are similar to those of LaAgSb$_2$,
FS 2 shows a complex three-dimensional geometry compared with the cylindrical shape in LaAgSb$_2$.
The obtained FS is consistent with the previous studies \cite{Hase_2014, Ruszala_2018}. 
Figure \ref{fig06}(b) shows the field-angular dependence of SdH frequencies calculated based on the FS shown
in Fig. \ref{fig06}(a).
Filled (open) markers represent the computational (experimental) results.
Around $\theta\sim0^{\circ}$, the calculation has no corresponding cross-sections. This suggests that the actual Fermi level slightly deviates from the calculation, presumably owing to unintentional doping caused by an imperfect stoichiometry.
We can assume that branches $\alpha$, $\beta$, and $\gamma$ are possible candidates
for the experimentally observed SdH oscillation
as they show similar $\theta$ dependencies with the experimental results.
The corresponding cross-sections for these branches are shown in Fig. \ref{fig06}(a),
and the expected $F$ and $m_c^*$ are listed in Table \ref{tab_qo}.
\begin{table}[]
\caption{\label{tab_qo}
Representative SdH frequency $F$ and cyclotron effective mass $m_c^*$ at $\theta=0^{\circ}$ in the case of $\epsilon-\epsilon_F = 0$ eV.
The values in the parenthesis indicate the case of $\epsilon-\epsilon_F = -0.16$ eV.
}
\begin{ruledtabular}
\begin{tabular}{lll}
label & $F$ (kT) & $m_c^*$ ($m_0$) \\
\hline
$\alpha$ & 0.027 & 0.22 \\
$\beta$ & 0.072 [0.33]& 0.16 [0.22]\\
$\gamma$ & 0.13 [0.050]& 0.065 [0.045]
\end{tabular}
\end{ruledtabular}
\end{table}
From the calculation, the $m_c^*$ values of $\alpha$ and $\beta$ are expected to be $0.22 m_0$ and $0.16 m_0$, respectively,
and only $\gamma$ can have $0.01 m_0$-order $m_c^*$ .
Thus, we attempted to shift the Fermi level within the rigid band approximation
to reproduce the experimental SdH data by the $\gamma$ branch.
The results are shown in Fig. \ref{fig06}(d),
and the corresponding FSs are shown in Fig. \ref{fig06}(c).
This required an energy shift of $\epsilon-\epsilon_F = -0.16$ eV,
which is shown by the horizontal solid line in Fig. \ref{fig05}.
As seen in Figs. \ref{fig06}(c) and \ref{fig06}(d), the hole (electron) FS gets fatter (thinner) by hole doping,
and an additional hole surface labeled FS 0 emerges.
This shift of $\epsilon_F$ corresponds to the change of approximately
0.2 electron and hole per unit cell, as described in Supplemental Material \cite{SM_URL}.
As shown in Table \ref{tab_qo},
the recalculated $F$ and $m_c^*$ of $\gamma$ branch are 50 T and $0.045 m_0$, respectively,
which shows reasonable agreement with the experimental value of 50 T and $0.052 m_0$.
From the results presented above, we can conclude that FS 4 is the most probable cause of the observed SdH oscillation.

\begin{figure}[]
\centering
\includegraphics[]{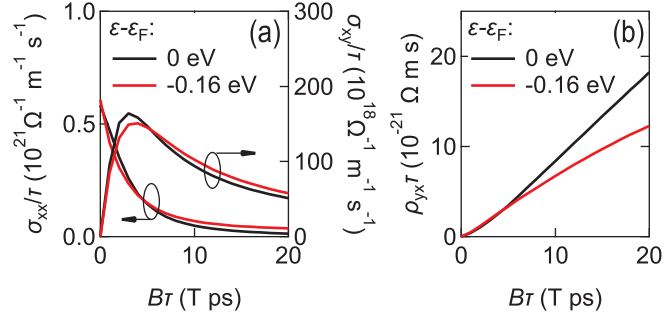}
\caption{
(a) Calculated conductivity tensors $\sigma_{xx, xy}/\tau$ as a function of $B\tau$ in the relaxation time approximation.
The curves shown in black (red) color represent undoped (hole-doped) cases.
(b) Calculated Hall resistivity $\rho_{yx}\tau=\sigma_{xy}\tau/(\sigma_{xx}^2+\sigma_{xy}^2)$.
The curves shown in black (red) color represent undoped (hole-doped) cases.
\label{fig07}}
\end{figure}

To gain more insights into the FS, we calculated the conductivity tensors and magnetic field dependence of Hall resistivity within the relaxation time ($\tau$) approximation.
Figure \ref{fig07}(a) represents the electrical conductivity and Hall conductivity divided by the relaxation time
($\sigma_{xx, xy}/\tau$) in the undoped ($\epsilon-\epsilon_F = 0$ eV) and
hole-doped ($\epsilon-\epsilon_F = -0.16$ eV) cases.
Herein, we assumed that $\tau$ is independent of the band index.
Based on these data, we can calculate the theoretical Hall response ($\rho_{yx}\tau=\sigma_{xy}\tau/(\sigma_{xx}^2+\sigma_{xy}^2)$) for each case, as shown in Fig. \ref{fig07}(b).
The calculation supports the fact that the Hall resistivity is positive,
which is consistent with the experimental results shown in Fig. \ref{fig03}(a).
Additionally, the nonlinearity of $\rho_{yx}\tau$ is more prominent in the hole-doped case, which qualitatively agrees with the experimental results.
Thus, our experimental results can be interpreted based on the slightly hole-doped FS shown in Fig. \ref{fig06}(c).

\begin{figure}[]
\centering
\includegraphics[]{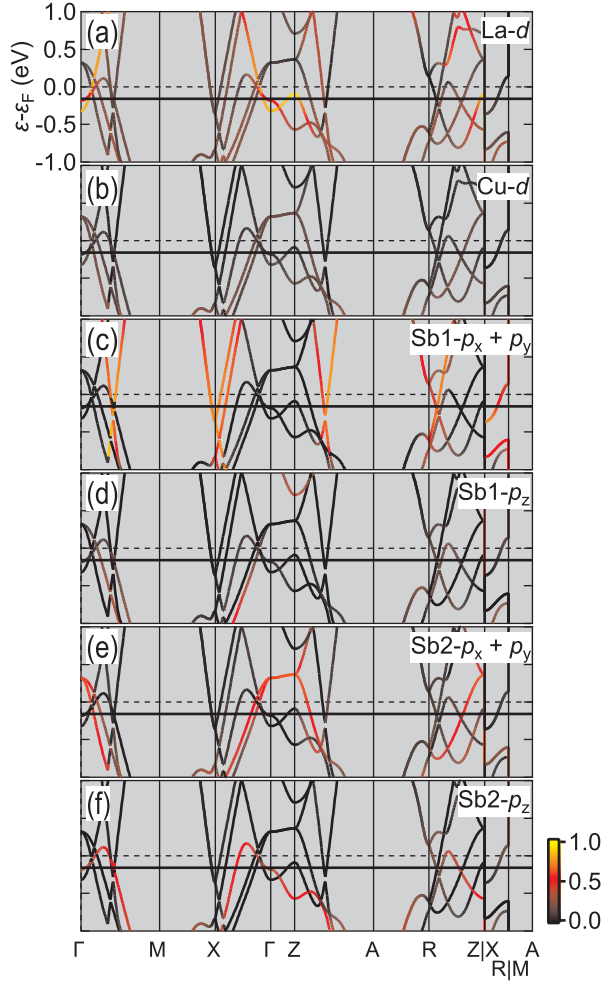}
\caption{
Orbital-projected band structure for (a) La-$d$, (b) Cu-$d$, (c) Sb1-$p_x+p_y$,
(d) Sb1-$p_z$, (e) Sb2-$p_x+p_y$, and (f) Sb2-$p_z$.
The color code indicates the ratio of the corresponding orbital characters.
\label{fig08}}
\end{figure}

Herein, we comment on the orbital character of the FS.
Figures \ref{fig08}(a)-(f) show the orbital projections on the band structure.
FS 3 and 4 have a Dirac-like steep dispersion
(around X and R point and on the way of $\Gamma$-M and Z-A path)
and consist of intensive $p_x$ and $p_y$-like characters of Sb1
with slight hybridization of La-$d$.
The emergence of this band crossing can be understood by the band folding of the $4^4$ square-net structure \cite{Hoffmann_1987, Klemenz_2019, Klemenz_2020}.
Owing to the steep dispersion,
the Fermi velocities $v_F$ of these FSs are higher than the others,
as indicated by the color code in Figs. \ref{fig06}(a) and \ref{fig06}(c).
Conversely, FS 0-2 consists of the orbitals of remaining atoms
(i.e., La-$d$, Cu-$d$, Sb2-$p$) and Sb1-$p_z$.
Among them, La-$d$ and Sb2-$p$ are particularly intensive,
while Cu-$d$ and Sb1-$p_z$ are relatively minor.
Considering the above, we can reasonably ascribe that the electron-like surfaces mainly derive from the Sb1 square-net layer,
and the hole-like surfaces from the interstitial structure between the Sb1-square nets in the real space.
An almost equivalent discussion applies in the case of LaAgSb$_2$,
whose orbital-projected band structure is shown in Supplemental Material \cite{SM_URL}.

As the position of the Fermi level has been identified above,
we can estimate the bare electron specific heat coefficient $\gamma_{bare}$
without any many-body effect using the calculated density of states.
Using a $D=3.0$ eV$^{-1}$ unit cell$^{-1}$ at $\epsilon-\epsilon_F = -0.16$ eV,
we obtained $\gamma_{bare}=3.5$ mJ mol$^{-1}$ K$^{-2}$.
The difference between the experimental $\gamma$ and $\gamma_{bare}$ should represent the mass enhancement achieved by the many-body effect, which in the present case was mainly derived from the electron--phonon interactions.
We defined the dimensionless EPC strength $\lambda$ as $\gamma = \gamma_{bare}(1+\lambda)$,
and obtained $\lambda = 0.33$.
The EPC should also cause the enhancement of $m_c^*$.
Using the bare cyclotron effective mass $m_c^*=0.045 m_0$ for $\gamma$ branch and $\lambda = 0.33$,
the enhanced mass should be $0.06 m_0$.
While the experimentally observed $m_c^*=0.052 m_0$ was slightly less than that expected, the general trend did not contradict the approximate estimation presented above.

\begin{figure*}[]
\centering
\includegraphics[]{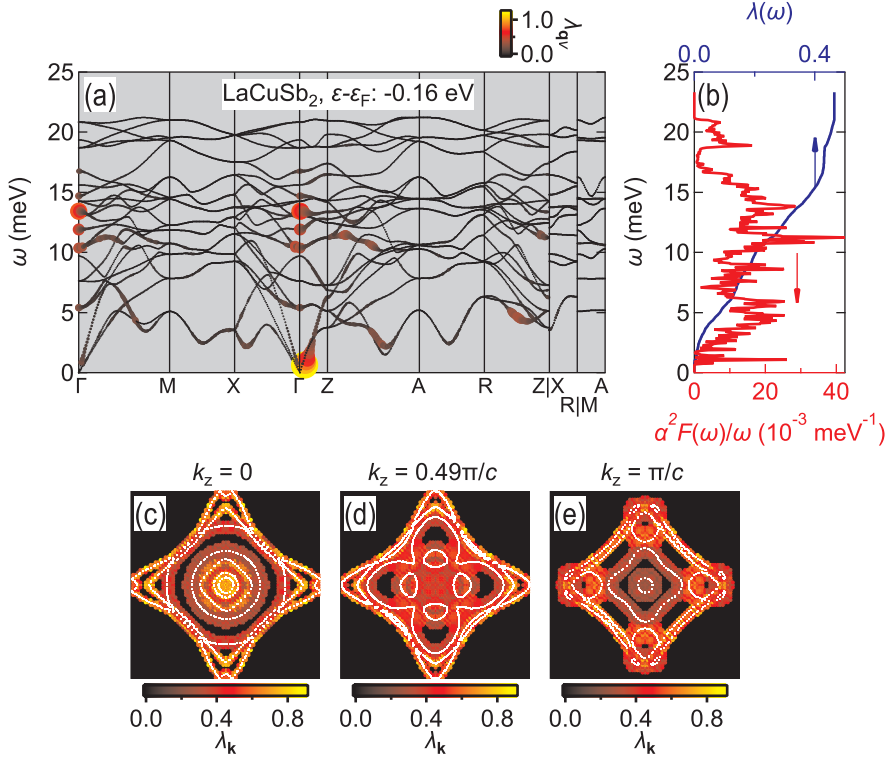}
\caption{
(a) Phonon dispersion and electron-phonon coupling strength $\lambda_{\bm{q}\nu}$ projected on the $\bm{q}$ space,
where $\bm{q}$ and $\nu$ represent the wavevector and mode index of corresponding phonon, respectively.
Color code and marker size indicate the magnitude of $\lambda_{\bm{q}\nu}$.
(b) Eliashberg spectral function divided by phonon frequency
$\alpha^2F(\omega)/\omega$ (bottom axis) and integrated electron--phonon coupling strength $\lambda(\omega)$ (top axis).
Electron--phonon coupling strength $\lambda_{\bm{k}}$ projected on the $k_{xy}$
plane at (c) $k_z=0$, (d) $k_z=0.49\pi/c$, and (e) $k_z=\pi/c$.
Color-coded and white markers indicate the magnitude of $\lambda_{\bm{k}}$ and cross-section of the FS, respectively.
\label{fig09}}
\end{figure*}

Subsequently, we discuss whether the observed SC can be explained by
conventional phonon-mediated mechanism.
Figure \ref{fig09}(a) shows the phonon dispersion of LaCuSb$_2$.
The phonon bands spread up to 22 meV,
and their energy scale is similar to that of LaAgSb$_2$.
In the present case, however,
we can see several phonon bands located at relatively low frequencies
around 5 meV,
which are not observed in the phonon dispersion of LaAgSb$_2$.
As shown in the phonon density of states in Supplemental Material \cite{SM_URL},
lattice vibrations related to Sb2 and Cu atoms are primarily responsible
for the low-frequency modes.

In the remaining parts, we focus on the EPC properties.
Herein, we performed the calculation
based on the hole-doped FS [Fig. \ref{fig06}(c)].
The color and radius of the markers
in Fig. \ref{fig09}(a) shows the EPC strength $\lambda_{\bm{q}\nu}$ for
each $\bm{q}$ and $\nu$.
For comparison, we also show an equivalent plot for LaAgSb$_2$ at ambient pressure in Supplemental Material \cite{SM_URL}.
Common to both cases, we can see the relatively strong EPC in optical phonons having energies in the range of 10--15 meV
near the zone center.
The intense EPC dominantly contributes to the Eliashberg spectral function divided by the phonon frequency $\alpha^2F(\omega)/\omega$, as shown in Fig. \ref{fig09}(b),
and thus results in the notable enhancement of integrated EPC strength $\lambda(\omega)$
[top axis in Fig. \ref{fig09}(b)]
in this energy region.

The important point to note in LaCuSb$_2$ is that the low-frequency phonon modes mentioned above show substantial EPC,
which results in a secondary peak structure in $\alpha^2F(\omega)/\omega$ at approximately 5 meV.
In Supplemental Material \cite{SM_URL},
we visualize the lattice vibrations at representative $\bm{q}$ points associated with intensive EPC.
It is shown that the low-frequency modes
(e.g., points D-G indicated in Supplemental Material \cite{SM_URL})
are relevant to atomic motions within the interstitial layer,
which is consistent with the atom-projected density of states.

The theoretical EPC strength was estimated to be $\lambda = 0.465$; this value is slightly higher but shows reasonable agreement with the experimental value of $\lambda = 0.33$.
The theoretical $\lambda$ value is approximately doubled compared with $\lambda=0.237$ in LaAgSb$_2$ at ambient pressure \cite{Akiba_2022_SC}.
In the case of LaAgSb$_2$,
the contribution to $\lambda$ up to 10 meV is only 0.044 (19 \% of the total $\lambda$).
By contrast, it reaches 0.21 (44 \%) in the case of LaCuSb$_2$.
We also note that the contribution to $\lambda$ above 10 meV is comparable in the LaCuSb$_2$ (0.26) and LaAgSb$_2$ (0.19) cases.
The results indicate that
the additional contribution from the low-frequency phonon modes
is crucial for the enhancement of $T_c$ in LaCuSb$_2$.

Figure \ref{fig09}(c)-(e) shows the cross-section of the FS
(white markers)
and distribution of the EPC strength
$\lambda_{\bm{k}}$
in the $\bm{k}$ space (color-coded markers).
Compared with the case of LaAgSb$_2$ shown in Supplemental Material \cite{SM_URL},
$\lambda_{\bm{k}}$ is less sensitive to the FS (i.e., more isotropic),
and the strength of $\lambda_{\bm{k}}$ is typically higher.
Notably, the magnitude of $\lambda_{\bm{k}}$ of the hole surfaces (located inside the hollow-shaped FS)
have comparable values to those of electron surfaces.
This is in contrast to the case of LaAgSb$_2$ in which only electron FSs with intense $p_x+p_y$ Sb1 characters have significant EPC.
The above supports the fact that not only Sb1 square-net layers but also the interstitial layers contribute to the superconducting properties in LaCuSb$_2$.

Finally, we deduced the theoretical $T_c$ value based on the McMilla--Allen--Dynes formalism to be $T_c^{MAD}=0.93$ K,
using $\lambda=0.465$, logarithmic average frequency $\omega_{log}=103.5$ K, and typical Coulomb pseudopotential $\mu^{*}=0.1$.
The correspondence with the experimental value ($T_c=1.0$ K) is quite reasonable.
Thus, we conclude that the SC of LaCuSb$_2$ derived from the conventional phonon-mediated mechanism, and seemed to be less involved with the criticality of a CDW order.

\section{Conclusions}
In conclusion, we investigated the electronic structure and superconducting properties of single-crystalline LaCuSb$_2$.
We certified by resistivity, magnetization, and specific heat measurements
that superconductivity is a bulk effect.
We observed the Shubnikov-de Haas oscillation at the frequency of 50 T and obtained an
effective cyclotron mass of 0.052 $m_0$,
which agreed with the findings of a previous study .
Contrary to the previous study, we observed monotonic field dependence of the magnetoresistivity and positive Hall resistivity.
We showed that the hole-doped condition,
which may be due to the imperfect stoichiometry,
explained the experimental results.
Based on the electronic structure determined above,
we investigated the electron-phonon coupling properties
to understand the superconductivity in LaCuSb$_2$.
The results clarify the difference from LaAgSb$_2$ that
(i) The vibration modes derived from the interstitial layer sandwiched between the Sb-square nets showed sizable electron-phonon coupling and
(ii) The momentum-resolved electron-phonon coupling distributed over the entire Fermi surface, i.e., all carriers contributed equally to the SC.
These facts are ascribed to be the origin of the enhanced superconducting transition temperature compared with LaAgSb$_2$.
Further, we showed that the theoretical superconducting transition temperature estimated by the McMillan--Allen--Dynes formula reasonably reproduced the experimental results.
Our study concludes that the SC of LaCuSb$_2$ can be understood within
the conventional framework of the phonon-mediated pairing mechanism.

\begin{acknowledgments}
This research was supported by JSPS KAKENHI Grants No. 19K14660, 21H01042, and 22K14006.
\end{acknowledgments}


\bibliography{reference}

\clearpage

\end{document}